# The Dark Side of ChatGPT: Legal and Ethical Challenges from Stochastic Parrots and Hallucination

Zihao Li[1,2]

[1] CREATe Centre, School of Law, University of Glasgow

[2] Stanford Law School, Stanford University

11 February 2023*

With the launch of ChatGPT, Large Language Models (LLMs) are shaking up our whole society, rapidly altering the way we think, create and live. For instance, the GPT integration in Bing has altered our approach to online searching. While nascent LLMs have many advantages, new legal and ethical risks are also emerging, stemming in particular from stochastic parrots and hallucination.[1] The EU is the first and foremost jurisdiction that has focused on the regulation of AI models. However, the risks posed by the new LLMs are likely to be underestimated by the emerging EU regulatory paradigm. Therefore, this correspondence warns that the European AI regulatory paradigm must evolve further to mitigate such risks.

## 1. Stochastic parrots and hallucination: unverified information generation

One potentially fatal flaw of the LLMs, exemplified by ChatGPT, is that the generation of information is unverified. For example, ChatGPT often generates pertinent, but non-existent academic reading lists. Data scientists claim that this problem is caused by "hallucination"[2] and "stochastic parrots".[3] Hallucination occurs when LLMs generate text based on their internal logic or patterns, rather than the true context,[2] leading to confidently but unjustified and unverified deceptive responses. Stochastic parrots is the repetition of training data or its patterns, rather than actual understanding or reasoning.

**The text production method of LLMs is to reuse, reshape, and recombine the training data in new ways to answer new questions while ignoring the problem of authenticity and trustworthiness of the answers.** In short, LLMs only predict the probability of a particular word coming next in a sequence, rather than actually comprehending its meaning. Although the majority of answers are high-quality and true, the content of the answers is fictional. Even though most training data is reliable and trustworthy, the essential issue is that the recombination of trustworthy data into new answers in a new context may lead to untrustworthiness, as the trustworthiness of information is conditional and often context-bound. If this precondition of trustworthy data disappears, trust in answers will be misplaced. Therefore, while the LLMs' answers may seem highly relevant to the prompts, they are made-up.

However, merely improving the accuracy of the models through new data and algorithms is insufficient, because the more accurate the model is, the more users will rely on it, and thus be tempted not to verify the answers, leading to greater risk when stochastic parrots and hallucinations appear. The risk is beyond measure if users encounter these problems in especially sensitive areas such as healthcare or the legal field. Even if utilizing real-time internet sources, the trustworthiness of LLMs may remain compromised, as exemplified by factual errors in new Bing's launch demo.

These risks can lead to ethical concerns, including misinformation and disinformation, which may adversely affect individuals through misunderstandings, erroneous decisions, loss of trust, and even physical harm (e.g., in healthcare). Misinformation and disinformation can reinforce bias,[4] as LLMs may perpetuate stereotypes present in their training data.[5]

## 2. The EU AI regulatory paradigm: Advanced Legal intervention required

The EU has already commenced putting effort into AI governance. The AI Act (AIA) is the first and globally most ambitious attempt to regulate AI. However, the proposed AIA, employing a risk-based taxonomy for AI regulation, encounters difficulties when applied to general-purpose LLMs. On the one hand, categorizing LLMs as high-risk AI due to its generality may impede EU AI development. On the other hand, if general-purpose LLMs are regarded as chatbots, falling within a limited-risk group, merely imposing transparency obligations (i.e., providers need to disclose that the answer is generated by AI) would be insufficient.[6] Because the danger of parroting and hallucination risks is not only related to whether users are clearly informed that they are interacting with AI, but also to the reliability and trustworthiness of LLMs' answers, i.e., how users can distinguish between truth and made-up answers. When a superficially eloquent and knowledgeable chatbot generates unverified content with apparent confidence, users may trust the fictitious content without undertaking verification. Therefore, the AIA's transparency obligation is not sufficient.

Additionally, the AIA does not address the role, rights, or responsibilities of the end-users. As a result, users have no chance to contest or complain about LLMs, especially when stochastic parrots and hallucination occur and affect their rights. Moreover, the AIA does not impose any obligations on users. However, as aforementioned, the occurrence of disinformation is largely due to deliberate misuse by users. Without imposing responsibilities on the user side, it is difficult to regulate the harmful use of AI by users.

Apart from the AIA, the Digital Service Act (DSA) aims to govern disinformation. However, the DSA's legislators only focus on the responsibilities of the intermediary, overlooking the source of the disinformation. Imposing obligations only on intermediaries when LLMs are embedded in services is insufficient, as such regulation cannot reach the underlying developers of LLMs. Similarly, the Digital Markets Act (DMA) focuses on the regulation of gatekeepers, aiming to establish a fair and competitive market. Although scholars recently claim that the DMA has significant implications for AI regulation,[7] the DMA primarily targets the effects of AI on market structure; it can only provide limited help on LLMs. The problem that the DSA and DMA will face is that both only govern the platform, not the usage, performance, and output of AI *per se*. This regulatory approach is a consequence of the current platform-as-a-service (PaaS) business model. However, once the business model shifts to AI model-as-a-service (MaaS),[8] this regulatory framework is likely to become nugatory, as the platform does not fully control the processing logic and output of the algorithmic model.

Therefore, it is necessary to urgently reconsider the regulation of general-purpose LLMs.[9] The parroting and hallucination issues show that minimal transparency obligations are insufficient, since LLMs often lull users into misplaced trust. When using LLMs, users should be acutely aware that the answers are made-up, may be unreliable, and require verification. LLMs should be obliged to remind and guide users on content verification. Particularly when prompted with sensitive topics, such as medical or legal inquiries, LLMs should refuse to answer, instead directing users to authoritative sources with traceable context. The suitable scope for such filter and notice obligations warrants further discussion from legal, ethical and technical standpoints.

Furthermore, legislators should reassess the risk-based AI taxonomy in the AIA. The above discussion suggests that the effective regulation of LLMs needs to ensure their trustworthiness, taking into account the reliability, explainability and traceability of generated information, rather than solely focusing on transparency. Meanwhile, end-users, developers and deployers' roles should all be considered in AI regulations, while shifting focus from PaaS to AI MaaS.